\documentclass[10pt]{article}

\newcommand{\be}{\begin{equation}}
\newcommand{\ee}{\end{equation}}

\def\b{\mbox{ }} % Blank tensor index

\newcommand{\Tr}{\mathop{\rm Tr}\nolimits}

\newcommand{\bfL}{{\bf L}}
\newcommand{\bfA}{{\bf A}}
\newcommand{\bfU}{{\bf U}}
\newcommand{\bfg}{{\bf g}}
\newcommand{\dbL}{\skew{-5}\dot{\bf L}}
\newcommand{\tbL}{\skew{-4}\tilde{\bf L}}
\newcommand{\bfLD}{\hat{\bf L}}
\newcommand{\bfAD}{\hat{\bf A}}
\newcommand{\bfBD}{\hat{\bf B}}
\newcommand{\bfGammaD}{\hat{\bf\Gamma}}

\let\case=\fraction  % for compatibility with ioplppt.sty

\catcode`@=11
\def\eqalign#1{\null\,\vcenter{\openup\jot\m@th
  \ialign{\strut\hfil$\displaystyle{##}$&$\displaystyle{{}##}$\hfil
      \crcr#1\crcr}}\,}
\def\meqalign#1{\null\,\vcenter{\openup\jot\m@th
  \ialign{\strut\hfil$\displaystyle{##}$&&$\displaystyle{{}##}$\hfil
      \crcr#1\crcr}}\,}
\catcode`@=12   % back to nonletter category 

\begin{document}

\bibliographystyle{unsrt}

\title{Lax pair tensors and integrable spacetimes}

\author{Kjell Rosquist \thanks{E-mail address: kr@vanosf.physto.se} \
  and Martin Goliath \thanks{E-mail address: goliath@vanosf.physto.se}}

\date{}

\maketitle

\vspace{-5mm}

\begin{center}
  Department of Physics, Stockholm University, Box 6730,\\
  S-113 85 Stockholm, Sweden
\end{center}

\begin{center}
  \sf gr-qc/9707003 \rm
\end{center}

\vspace{5mm}

\begin{abstract}
  The use of Lax pair tensors as a unifying framework for Killing tensors
  of arbitrary rank is discussed. Some properties of the
  tensorial Lax pair formulation are stated. A mechanical system with a
  well-known Lax representation -- the three-particle open Toda lattice -- is
  geometrized by a suitable canonical transformation. In this way the
  Toda lattice is realized as the geodesic system of a certain
  Riemannian geometry. By using different canonical transformations we
  obtain two inequivalent geometries which both represent the
  original system. Adding a timelike dimension gives four-dimensional
  spacetimes which admit two Killing vector fields and are completely
  integrable. 
\end{abstract}

\section{Introduction}
Many problems in general relativity require an understanding of the global
structure of the spacetime. Currently discussed global problems include the
occurrence of naked singularities \cite{art:OriPiran1987} and
universality in gravitational collapse situations
\cite{art:Choptuik1993}. The study of global properties of spacetimes 
relies to a large extent on the ability to integrate the geodesic
equations. In the absence of exact solutions numerical integration is often
used to obtain a quantitative picture. However, in the quest for a deeper
understanding the exact and numerical approaches should be viewed as
complementary tools. To perform an exact investigation of the global
properties of a given spacetime, not only must the spacetime itself be an exact
solution of the Einstein equations, but in addition the geodesic equations
must be integrable. Usually, in a $d$-dimensional space, integrability
of the geodesic equations is 
connected with the existence of at least $d-1$ mutually commuting Killing
vector fields which span a hypersurface in the spacetime. There are exceptions
however. The most well-known example is the Kerr spacetime which has only two
commuting Killing vectors. In that case it is the existence of an irreducible
second rank Killing tensor which makes integration possible
\cite{art:WalkerPenrose1970}. Another 
example is given by Ozsvath's class III cosmologies
\cite{art:Ozsvath1970}.
In that case the geodesic
system was integrated using the existence of a non-abelian Lie algebra of four
Killing vectors \cite{art:Rosquist1980}. In general integrability can
only be guaranteed if there is a 
set of $d$ constants of the motion in involution ({\em i.e.} mutually Poisson
commuting). Since the metric itself always provides one constant of the motion
corresponding to the squared length of the geodesic tangent vector, the
geodesic system will be integrable by Liouville's theorem if there are $d-1$
additional Poisson commuting invariants.

Exact solutions of Einstein's equations typically admit a number of Killing
vector fields. Some of these Killing vector fields may be motivated by
physical considerations. For example if one is interested in static stars the
spacetime must have a timelike Killing vector. For such systems it is also
very reasonable to assume spherical spatial symmetry leading to a total of
four (noncommuting) Killing vectors. In most cases, the number of Killing
vectors is limited by the physics of the problem. In a spherically symmetric
collapse situation, for example, the spacetime admits exactly three
noncommuting Killing vector fields which form an isometry group with
2-dimensional orbits. That structure is not sufficient for an exact
integration of the geodesic equations. However, the physics of the problem
does not impose any {\em a priori\/} restrictions on higher rank ($\ge2$)
Killing tensors. A Killing vector field, $\xi$, plays a double role; it is
both an isometry for the metric (${\cal L}_\xi g =0$) and a geodesic
symmetry. This last property means that it can be interpreted as a symmetry
transformation for the geodesic equations. By contrast higher rank Killing
tensors are only geodesic symmetries. They have no obvious geometric
interpretation (but {\em cf.}  \cite{art:Rosquist1989}). Because of the
isometry property of the Killing vector fields, such symmetries can be
incorporated right from the start by assuming a particular form the metric. In
this way the field equations are actually simplified by the assumption of
Killing vector symmetries. On the other hand, the higher rank Killing
symmetries can at present not be used to simplify the form of the field
equations. Instead the Killing tensor equations must be imposed as extra
conditions thereby increasing both the number of dependent variables and the
number of equations.

The Lax tensors introduced in \cite{art:Rosquist1997} provide a unifying
framework for Killing tensors of any rank and may lead to possibilities to
incorporate the higher Killing symmetries in the field equations
themselves. We will comment briefly on this issue below. A single Lax tensor
may generate Killing tensors of varying ranks. Lax tensors arise from a
covariant formulation of the Lax pair equation \cite{art:Lax1968} for
Riemannian and pseudo-Riemannian geometries. The standard Lax pair formulation
involves a pair of matrices. In the covariant formulation on the other hand,
the Lax pair is represented by two third rank tensors. The first Lax matrix
corresponds exactly to the first Lax tensor while the second Lax matrix and
the second Lax tensor differ by a term which coincides with the Levi-Civita
connection. The derivative part of the tensorial Lax pair equation is
identical to the Killing-Yano equation. Therefore Killing-Yano tensors are
special cases of Lax tensors for which the second Lax tensor vanishes (the
second Lax matrix however does not). However, whereas Killing-Yano tensors are
by definition totally antisymmetric the Lax tensors have no {\em a priori\/}
symmetry restrictions.  In this paper we discuss methods for constructing
spacetimes which admit a nontrivial pair of Lax tensors. We also give two
examples of such spacetimes.

\section{Lax pair tensors}
In this section we outline the approach to integrable geometries as
given in \cite{art:Rosquist1997}. We consider a Riemannian or pseudo-Riemannian
geometry with metric
\be
ds^2 = g_{\mu\nu} dq^\mu dq^\nu \ .   
\ee
The geodesic equations can be represented by the Hamiltonian
\be\label{eq:Hdef} 
H = \case12 g^{\mu\nu} p_\mu p_\nu \ ,
\ee
together with the natural Poisson bracket (denoted by $\{,\}$) on the
cotangent bundle. The geodesic system is given by
\be
\dot q^\alpha = \{q^\alpha, H\} = g^{\alpha\mu} p_\mu \ ,\qquad
\dot p_\alpha = \{p_\alpha, H\} = \Gamma^{\mu\nu}{}_\alpha \,
p_\mu p_\nu .
\ee 
The complete integrability of this system can be shown
with the help of a pair of matrices $\bfL$ and $\bfA$ with entries
defined on the phase space (the cotangent bundle) and satisfying the
Lax pair equation \cite{art:Lax1968}
\be\label{eq:lax}
\dbL = \{\bfL, H\} = [\bfL, \bfA] \ .
\ee
It follows from (\ref{eq:lax}) that the quantities $I_k := \case1k
\Tr\bfL^k$ are all constants of the motion. If in addition they commute with
each other $\{I_k, I_j\}=0$ (Liouville integrability) then it is possible to
integrate the system completely at least in principle
(see e.g. \cite{book:Arnold1989}).
The Lax representation (\ref{eq:lax}) is not unique. In fact, the Lax
pair equation is invariant under a transformation of the form
\be\label{eq:simitransf}
\tbL = \bfU\bfL\bfU^{-1} \ ,\qquad
\tilde\bfA = \bfU\bfA\bfU^{-1} - \dot\bfU\bfU^{-1} \ .
\ee
We see that $\bfL$ transforms as a tensor while $\bfA$ transforms as a
connection. As we will see, these statements acquire a more precise
meaning in the geometric formulation which we will now describe.\\
Typically, the Lax matrices are linear in the momenta and in the geometric
setting they may also be assumed to be homogeneous. This motivates the
introduction of two third rank geometrical objects,
$L^\alpha{}_\beta{}^\gamma$ and  $A^\alpha{}_\beta{}^\gamma$, such that the
Lax matrices can be written in the form \cite{art:Rosquist1997}
\be
\bfL = (L^\alpha{}_\beta) = (L^\alpha{}_\beta{}^\mu p_\mu) \ ,\qquad
\bfA = (A^\alpha{}_\beta) = (A^\alpha{}_\beta{}^\mu p_\mu) \ .
\ee
We will refer to $L^\alpha{}_\beta{}^\gamma$ and
$A^\alpha{}_\beta{}^\gamma$ as the Lax tensor and the Lax connection,
respectively. Defining
\be
{\bf B} = (B^\alpha_{\b \beta}) = (B^\alpha{}_\beta{}^\mu p_\mu) =
{\bf A} - {\bf \Gamma} ,
\ee
where
\be
{\bf \Gamma} = (\Gamma^\alpha_{\b \beta}) =
(\Gamma^\alpha{}_\beta{}^\mu p_\mu)
\ee
is the Levi-Civita connection with respect to $g_{\alpha\beta}$, it
then follows that the Lax pair equation takes the covariant form (see
\cite{art:Rosquist1997} for details) 
\be
L^\alpha{}_{\beta}{}^{(\gamma;\delta)} = 
  L^\alpha{}_\mu{}^{(\gamma} B^{|\mu|}{}_{\beta}{}^{\delta)}
- B^\alpha{}_\mu{}^{(\gamma} L^{|\mu|}{}_\beta{}^{\delta)} \ ,
\ee
where $L^\alpha{}_\beta{}^\gamma$ and $B^\alpha{}_\beta{}^\gamma$ are
tensorial objects. Note that the right-hand side of this equation is
traceless, so that upon contracting over $\alpha$ and $\beta$ we
obtain the Killing vector equation $L^\mu{}_{\mu(\alpha;\beta)} = 0$.
Splitting the Lax tensors in symmetric 
and antisymmetric parts with respect to the first two indices,
     $S_{\alpha\beta\gamma} = L_{(\alpha\beta)\gamma}$, 
$R_{\alpha\beta\gamma} = L_{[\alpha\beta]\gamma}$ and $P_{\alpha\beta\gamma}
= B_{(\alpha\beta)\gamma}$, $Q_{\alpha\beta\gamma} =
B_{[\alpha\beta]\gamma}$, the Lax pair equation can be written as the system
\be\eqalign{
     S^{\alpha\beta}{}_{(\gamma;\delta)} &= 
      -2 S^{(\alpha}{}_{\mu(\gamma} Q^{\beta)\mu}{}_{\delta)}
      +2 R^{(\alpha}{}_{\mu(\gamma} P^{\beta)\mu}{}_{\delta)}  \ ,\cr
     R^{\alpha\beta}{}_{(\gamma;\delta)} &=
      -2 R^{[\alpha}{}_{\mu(\gamma} Q^{\beta]\mu}{}_{\delta)}
      +2 S^{[\alpha}{}_{\mu(\gamma} P^{\beta]\mu}{}_{\delta)} \ .\cr
}\ee
It is evident that this system is coupled via $P_{\alpha\beta\gamma}$. 
Setting $P_{\alpha\beta\gamma} = 0$ gives the two separate sets of equations
\be\label{eq:slax}
S^{\alpha\beta}{}_{(\gamma;\delta)} = 
-2 S^{(\alpha}{}_{\mu(\gamma} Q^{\beta)\mu}{}_{\delta)}  \ ,
\ee
\be\label{eq:alax}
R^{\alpha\beta}{}_{(\gamma;\delta)} =
-2 R^{[\alpha}{}_{\mu(\gamma} Q^{\beta]\mu}{}_{\delta)} \ .
\ee
We will see below that the Lax tensors $L_{\alpha\beta\gamma}$ and
$B_{\alpha\beta\gamma}$ in a geometrized version of the open Toda lattice 
are symmetric and antisymmetric respectively and therefore satisfy
(\ref{eq:slax}). If $R_{\alpha\beta\gamma}$ is totally antisymmetric
(with respect to all three indices) and $Q_{\alpha\beta\gamma} =0$, then the
equations (\ref{eq:alax}) are identical to the third rank Killing-Yano
equations \cite{art:Yano1952}. Therefore third rank Killing-Yano
tensors are special cases of Lax tensors. 

It is possible but not necessary to identify the invariant $I_2$ with the
geodesic Hamiltonian (\ref{eq:Hdef}). If such an identification is done then
the metric is given by the relation
\be
     g^{\alpha\beta} = L^\mu{}_\nu{}^\alpha L^\nu{}_\mu{}^\beta \ .  
\ee
Defining matrices $\bfL^\mu$ with components $(\bfL^\mu)^\alpha{}_\beta =
L^\alpha{}_\beta{}^\mu$, the metric components are given by the formula
\be
    g^{\alpha\beta} = \Tr(\bfL^\alpha\bfL^\beta) \ .
\ee
This formula suggests using the components of the $\bfL^\mu$ (or some internal
variables from which the $\bfL^\mu$ are built) as the basic variables already
in the formulation of the field equations much like in the Ashtekar variable
formalism \cite{book:Ashtekar1988}.

\section{Geometrization and tensorial representations of the
     three-particle open Toda lattice}
Integrable systems are usually discussed in the context of classical
mechanics. Classical Hamiltonians typically consist of a flat
positive-definite kinetic energy together with a potential energy
term. They are thus superficially quite different from geometric
Hamiltonians of the form (\ref{eq:Hdef}). However, any classical
Hamiltonian with a quadratic kinetic energy can be transformed to a
geometric representation. One such geometrization results in the
Jacobi Hamiltonian \cite{book:Lanczos1986}. Another closely related
geometrization was used in \cite{art:Rosquist1997}. Both methods
involve a reparameterization of the independent variable. Usually we
will refer to the independent variable as the time, although its
physical interpretation may vary. As a
consequence of this feature, the original Lax representation is not
preserved. It is known how to transform the invariants themselves
under the time reparameterization
\cite{art:RosquistPucacco1995,art:Rosquist1997}. Given that the
geometrized invariants are also in involution, the existence of a Lax
representation is guaranteed \cite{art:BabelonViallet1990}. However,
to actually find such a Lax representation is non-trivial. Another
geometrization scheme which does preserve the original Lax
representation is to apply a suitable canonical transformation. This
is however only possible for Hamiltonians with a potential of a
special form. One such system that we will consider in this paper is
the three-particle open Toda lattice 
\be\label{eq:hamil}
H = \case12 \left({\bar p}_1{}^2 + {\bar p}_2{}^2 + {\bar p}_3{}^2 \right) + 
e^{2({\bar q}^1 - {\bar q}^2)} + e^{2({\bar q}^2 - {\bar q}^3)}.
\ee
Below we will discuss two canonical transformations which
correspond to inequivalent geometric representations of
(\ref{eq:hamil}). For an explicit 
integration of the Toda lattice, see e.g. \cite{book:Perelomov1990}.
The standard symmetric Lax representation is \cite{book:Perelomov1990}
\be
\bfL = \left( 
\begin{array}{ccc}
  {\bar p}_1 & {\bar a}_1 & 0  \\
  {\bar a}_1 & {\bar p}_2 & {\bar a}_2 \\
  0 & {\bar a}_2 & {\bar p}_3
\end{array} \right),\hspace{1cm}
\bfA = \left( 
\begin{array}{ccc}
  0 & {\bar a}_1 & 0  \\
  -{\bar a}_1 & 0 & {\bar a}_2 \\
  0 & -{\bar a}_2 & 0
\end{array} \right),
\ee
where
\be
\eqalign{
{\bar a}_1 &= \exp({\bar q}^1 - {\bar q}^2) , \cr
{\bar a}_2 &= \exp({\bar q}^2 - {\bar q}^3) .
}\ee
Note that the definitions of ${\bar a}_1$ and ${\bar a}_2$ differ from
the ones used in \cite{art:Rosquist1997}.
The Hamiltonian (\ref{eq:hamil}) admits the linear invariant $I_1 = \Tr \bfL
= {\bar p}_1 + {\bar p}_2 + {\bar p}_3$, corresponding to
translational invariance. The Lax representation also gives rise to
the two invariants $I_2 = \case12 \Tr \bfL^2 = H$ and $I_3 = \case13
\Tr \bfL^3$.
As discussed above, we will assume that the tensorial Lax
representation is linear and homogeneous in the momenta. A homogeneous
Lax representation can be obtained from the standard
representation by applying a canonical transformation of the phase
space. This can be done in several ways. We will investigate two
possibilities below.

\subsection{Tensorial Lax representation I}
In the first attempt we straightforwardly apply a simple canonical
transformation that will give a linear and homogeneous Lax representation
\be\meqalign{
  {\bar q}^1 &= q^1 + \ln p_1,  \hspace{1cm} &{\bar p}_1 = p_1 , \cr
  {\bar q}^2 &= q^2,  &{\bar p}_2 = p_2  , \cr
  {\bar q}^3 &= q^3 - \ln p_3,  &{\bar p}_3 = p_3 . 
}\ee
The resulting Lax pair matrices are
\be\label{eq:repr1}
\bfL = \left( 
\begin{array}{ccc}
  p_1 & a_1 \, p_1 & 0  \\
  a_1 \, p_1 & p_2 & a_2 \, p_3 \\
  0 & a_2 \, p_3 & p_3
\end{array} \right),\hspace{1cm}
\bfA = \left( 
\begin{array}{ccc}
  0 & a_1 \, p_1 & 0  \\
  -a_1 \, p_1 & 0 & a_2 \, p_3 \\
  0 & -a_2 \, p_3 & 0
\end{array} \right),
\ee
where
\be\eqalign{
  a_1 &= \exp(q^1 - q^2), \cr
  a_2 &= \exp(q^2 - q^3).
}\ee
The Hamiltonian is now purely kinetic
\be
H = \case12 \Tr \bfL^2 = \case12 \left[ \left(1 + 2 a_1{}^2 \right)
p_1{}^2 + p_2{}^2 +  \left(1 + 2 a_2{}^2 \right) p_3{}^2 \right] .
\ee
Using (\ref{eq:Hdef}) we identify a metric
\be\label{eq:metric1}
ds^2 = g_{11} (dq^1)^2 + (dq^2)^2 + g_{33} (dq^3)^2 ,
\ee
where
\be\eqalign{
  g_{11} &= \left(1 + 2 a_1{}^2 \right)^{-1}, \cr
  g_{33} &= \left(1 + 2 a_2{}^2 \right)^{-1}.
}\ee
The non-zero Levi-Civita connection coefficients,
$\Gamma_{\b \beta\gamma}^\alpha = \Gamma_{\b (\beta\gamma)}^\alpha$, of
this metric are
\be\meqalign{
  \Gamma^1_{\b 11} &= -2 a_1{}^2 g_{11} , \hspace{10mm}
  & \Gamma^2_{\b 33} = 2 a_2{}^2 (g_{33})^2 , \cr
  \Gamma^1_{\b 12} &= 2 a_1{}^2 g_{11} ,
  & \Gamma^3_{\b 23} = -2 a_2{}^2 g_{33} , \cr
  \Gamma^2_{\b 11} &= -2 a_1{}^2 (g_{11})^2 ,
  & \Gamma^3_{\b 33} = 2 a_2{}^2 g_{33} . 
}\ee
Following the arguments above, the homogeneous Lax matrix
should correspond to a tensor with mixed indices $L^\alpha_{\b \beta}$.
It is a reasonable assumption that the covariant Lax formulation
inherits the symmetries of the standard formulation we started with.
We therefore expect $L_{\alpha\beta}$ and $B_{\alpha\beta}$ to have
the symmetries $L_{(\alpha\beta)} = L_{\alpha\beta}$ and
$B_{[\alpha\beta]} = B_{\alpha\beta}$. Note that the symmetry
properties are not imposed on the Lax matrices, $L^\alpha_{\b
  \beta}$ and $B^\alpha_{\b \beta}$, themselves. In fact, the required
symmetries are not consistent with the representation
(\ref{eq:repr1}). We can however perform a similarity 
transformation (\ref{eq:simitransf}) of the Lax matrix, $\bfL
\rightarrow \tbL$ in such a way that $\tilde L_{\alpha\beta}$ will be
symmetric. Using the transformation matrix
\be
\bfU = \left( 
\begin{array}{ccc}
  1/\sqrt{g_{11}} & 0 & 0  \\
  0 & 1 & 0 \\
  0 & 0 & 1/\sqrt{g_{33}}
\end{array} \right)
\ee
will give a new Lax pair 
\be
\bfL = \left( 
\begin{array}{ccc}
  p_1 & a_1 /\sqrt{g_{11}} \, p_1 & 0  \\
  a_1 \sqrt{g_{11}} \, p_1 & p_2 & a_2 \sqrt{g_{33}} \, p_3 \\
  0 & a_2 / \sqrt{g_{33}} \, p_3 & p_3
\end{array} \right),
\ee
\be  
\bfA = \left( 
\begin{array}{ccc}
  \Gamma^{1 \b 1}_{\b 1} \, p_1 + \Gamma^{1 \b 2}_{\b 1} \, p_2 & 
  a_1 /\sqrt{g_{11}} \, p_1 & 0  \\
  -a_1 \sqrt{g_{11}} \, p_1 & 0 & a_2 \sqrt{g_{33}} \, p_3 \\
  0 & -a_2 / \sqrt{g_{33}} \, p_3 & \Gamma^{3 \b 2}_{\b 3} \, p_2 +
  \Gamma^{3 \b 3}_{\b 3} \, p_3
\end{array} \right),
\ee
where $\bfL$ is such that $L_{\alpha\beta}$ is symmetric. Defining
$\bfLD = (L_{\alpha\beta})$ and $\bfAD = (A_{\alpha\beta})$ by
\be
\bfLD = \bfg \bfL \ ,\qquad   \bfAD = \bfg \bfA \ ,
\ee
where $\bfg = (g_{\alpha\beta})$, we have
\be
\bfLD = \pmatrix{{g_{11}p_1}&{a_1\sqrt {g_{11}}\,p_1}&0\cr
  {a_1\sqrt {g_{11}}\,p_1}&{p_2}&{a_2\sqrt {g_{33}}\,p_3}\cr
  0&{a_2\sqrt {g_{33}}\,p_3}&{g_{33}p_3}\cr } ,
\ee
and 
\be
\bfAD = \pmatrix{
  \Gamma_{11}{}^1 p_1+\Gamma_{11}{}^2 p_2&{a_1\sqrt{g_{11}}\,p_1}&0\cr
  {-a_1\sqrt {g_{11}}\,p_1}& 0 &{a_2\sqrt {g_{33}}\,p_3}\cr
  0&{-a_2\sqrt {g_{33}}\,p_3}&\Gamma_{33}{}^2 p_2+\Gamma_{33}{}^3 p_3\cr} .
\ee
Note that the upper triangular parts of $\bfLD$ and $\bfAD$ coincide. This
property is peculiar to the open Toda lattice. 
We also define the corresponding connection matrix
$\bfGammaD = {\bf g \Gamma}$ given by
\be
\bfGammaD = \pmatrix{
  {\Gamma_{11}{}^1p_1+\Gamma_{11}{}^2p_2}&{2a_1{}^2g_{11}p_1}&0\cr
  {-2a_1{}^2g_{11}p_1}&0&{2a_2{}^2g_{33}p_3}\cr
  0&{-2a_2{}^2g_{33}p_3}&{\Gamma_{33}{}^2p_2+\Gamma_{33}{}^3p_3}\cr} \ .
\ee
We see that the off-diagonal part of the matrix $\bfGammaD$ is antisymmetric
like that of $\bfAD$ and furthermore that their off-diagonal components are
related by the simple relation $\Gamma_{\alpha\beta}{}^\gamma =
2(A_{\alpha\beta}{}^\gamma)^2$ for $\alpha < \beta$. This
gives the following relation between the upper triangular parts of $\bfLD$
and $\bfAD$
\be\label{eq:LBGrel}
\Gamma_{\alpha\beta}{}^\gamma = 2(L_{\alpha\beta}{}^\gamma)^2 \ ,
\quad ({\rm for\ }\alpha<\beta) \ .
\ee
Using the relation $\bfAD = \bfGammaD + \bfBD$ where
$\bfBD = {\bf g B}$ we then find the
following relation between the upper triangular 
components of $\bfLD$ and $\bfBD$
\be
B_{\alpha\beta}{}^\gamma = L_{\alpha\beta}{}^\gamma -
2(L_{\alpha\beta}{}^\gamma)^2 \ , \quad ({\rm for\ }\alpha<\beta) \ .
\ee
Finally expressing $\bfLD$ and $\bfBD$ in terms of $\bfGammaD$ we have for
the upper triangular parts
\be
L_{\alpha\beta}{}^\gamma = \sqrt{\case12\Gamma_{\alpha\beta}{}^\gamma}
\ ,\qquad
B_{\alpha\beta}{}^\gamma = -\Gamma_{\alpha\beta}{}^\gamma +
\sqrt{\case12\Gamma_{\alpha\beta}{}^\gamma}\ ,\qquad
({\rm for\ }\alpha<\beta) \ .
\ee
Furthermore, the diagonal elements of $\bfAD$ and $\bfGammaD$ are
identical. This implies that $\bfBD$ is antisymmetric in
agreement with our expectations.

\subsection{Tensorial Lax representation II}
The canonical transformation used in the previous section is not the
 only possible choice. The Toda lattice (\ref{eq:hamil}) has a Killing vector
symmetry. In fact, by adapting the coordinates $(\bar{q}^i,
\bar{p}_i)$ to the linear symmetry, another representation is suggested.
For this purpose, a suitable canonical transformation is
\be\meqalign{
  & \bar{q}^1 = \case1{\sqrt{2}} \tilde{q}^1 + \case1{\sqrt{6}}
  \tilde{q}^2 + \case1{\sqrt{3}} \tilde{q}^3 , 
  & \bar{p}_1 = \case1{\sqrt{2}} \tilde{p}_1 + \case1{\sqrt{6}}
  \tilde{p}_2 + \case1{\sqrt{3}} \tilde{p}_3, \cr
  & \bar{q}^2 = -\sqrt{\case23} \tilde{q}^2 +
  \case1{\sqrt{3}} \tilde{q}^3 , 
  & \bar{p}_2 = -\sqrt{\case23} \tilde{p}_2 +
  \case1{\sqrt{3}} \tilde{p}_3,  \cr
  & \bar{q}^3 = -\case1{\sqrt{2}} \tilde{q}^1 + \case1{\sqrt{6}}
  \tilde{q}^2 + \case1{\sqrt{3}} \tilde{q}^3, \hspace{5mm} 
  & \bar{p}_3 = -\case1{\sqrt{2}} \tilde{p}_1 + \case1{\sqrt{6}}
  \tilde{p}_2 + \case1{\sqrt{3}} \tilde{p}_3 .
}\ee
The Hamiltonian then becomes
\be
H = \case12 \left({\tilde p}_1{}^2 + {\tilde p}_2{}^2 +
{\tilde p}_3{}^2 \right) + 
 2 \, e^{\sqrt{2} {\tilde q}^1} \, \cosh(\sqrt{6} {\tilde q}^2) .
\ee
The linear invariant is $I_1 = \sqrt{3} {\tilde p}_3$. The form of the
Hamiltonian now suggests applying a canonical transformation of the form
\be\meqalign{
  & {\tilde q}^1 = q^1 + \sqrt{2} \ln p_1, \hspace{1cm} 
  & {\tilde p}_1 = p_1, \cr
  & {\tilde q}^2 = q^2, 
  & {\tilde p}_2 = p_2, \cr
  & {\tilde q}^3 = q^3 , 
  & {\tilde p}_3 = p_3 .
}\ee
The resulting homogeneous Lax representation is 
\be
\bfL = \left( 
\begin{array}{ccc}
  {1 \over \sqrt{2}} p_1 + {1 \over \sqrt{6}} p_2  &
  a_1 p_1 & 0  \\
  a_1 p_1 & -\sqrt {2 \over 3} p_2 & a_2 p_1 \\
  0 & a_2 p_1 & 
  -{1 \over \sqrt{2}} p_1 + {1 \over \sqrt{6}} p_2  
\end{array} \right) + \case1{\sqrt{3}} p_3 \, {\bf 1} ,
\ee
\be
\bfA = \left( 
\begin{array}{ccc}
  0 & a_1 p_1 & 0  \\
  -a_1 p_1 & 0 & a_2 p_1 \\
  0 & -a_2 p_1 & 0
\end{array} \right) ,
\ee
where 
\be\eqalign{
  a_1 &= \exp(\case1{\sqrt{2}} q^1 + \sqrt{\case32} q^2) , \cr
  a_2 &= \exp(\case1{\sqrt{2}} q^1 - \sqrt{\case32} q^2) .
}\ee
This gives the purely kinetic Hamiltonian 
\be
H = \case12 \Tr \bfL^2 = \case12 \left\{ \left[ 1 + 2 (a_1{}^2 + a_2{}^2) \right]
p_1{}^2 + p_2{}^2 +  p_3{}^2 \right\} ,
\ee
and the corresponding metric becomes
\be\label{eq:metric2}
ds^2 = g_{11} (dq^1)^2 + (dq^2)^2 + (dq^3)^2 ,
\ee
where 
\be
g_{11} = \left[ 1 + 2(a_1{}^2 + a_2{}^2) \right]^{-1} .
\ee
The non-zero Levi-Civita connection coefficients of this metric are
\be\eqalign{
  \Gamma^1_{\b 11} &= -\sqrt{2} (a_1{}^2 + a_2{}^2) g_{11} , \cr
  \Gamma^1_{\b 12} &= -\sqrt{6} (a_1{}^2 - a_2{}^2) g_{11} , \cr
  \Gamma^2_{\b 11} &=  \sqrt{6} (a_1{}^2 - a_2{}^2) (g_{11})^2 .
}\ee
Making a similarity transformation (\ref{eq:simitransf}) with
\be
\bfU = \left( 
\begin{array}{ccc}
  1/\sqrt{g_{11}} & 0 & 0  \\
  0 & 1 & 0 \\
  0 & 0 & 1
\end{array} \right)
\ee
we obtain a Lax pair with the desired symmetry properties
\be
\bfL = \left( 
\begin{array}{ccc}
  {1 \over \sqrt{2}} p_1 + {1 \over \sqrt{6}} p_2 &
  a_1 / \sqrt{g_{11}} p_1 & 0  \\
  a_1 \sqrt{g_{11}} p_1 & -\sqrt {2 \over 3} p_2 & a_2 p_1 \\
  0 & a_2 p_1 & -{1 \over \sqrt{2}} p_1 + {1 \over \sqrt{6}} p_2
\end{array} \right)  + \case1{\sqrt{3}} p_3 \, {\bf 1} ,
\ee
\be
\bfA = \left( 
\begin{array}{ccc}
  \Gamma^{1 \b 1}_{\b 1} \, p_1 + \Gamma^{1 \b 2}_{\b 1} \, p_2 & 
  a_1 / \sqrt{g_{11}} p_1 & 0  \\
  -a_1 \sqrt{g_{11}} p_1 & 0 & a_2 p_1 \\
  0 & -a_2 p_1 & 0
\end{array} \right) .
\ee
The matrices $\bfLD$ and $\bfAD$ have the form
\be
\bfLD = \left( 
\begin{array}{ccc}
  \left({1 \over \sqrt{2}} p_1 + {1 \over \sqrt{6}} p_2 \right) g_{11} &
    a_1 \sqrt{g_{11}} p_1 & 0  \\
    a_1 \sqrt{g_{11}} p_1 & -\sqrt {2 \over 3} p_2 & a_2 p_1 \\
    0 & a_2 p_1 & -{1 \over \sqrt{2}} p_1 + {1 \over \sqrt{6}} p_2 
  \end{array} \right) + \case1{\sqrt{3}} p_3 \, {\bf g} ,
\ee
\be
\bfAD = \left( 
\begin{array}{ccc}
  \Gamma_{11}{}^1 \, p_1 + \Gamma_{11}{}^2 \, p_2 & 
  a_1 \sqrt{g_{11}} p_1 & 0  \\
  -a_1 \sqrt{g_{11}} p_1 & 0 & a_2 p_1 \\
  0 & -a_2 p_1 & 0
\end{array} \right) .
\ee
As in the first case, the upper triangular parts of $\bfLD$ and
$\bfAD$ coincide. There is however no simple relation like
(\ref{eq:LBGrel}) between the components of $\bfLD$ and the
corresponding components of the connection matrix $\bfGammaD$. The form of
${\bf {\hat B}} = \bfAD - \bfGammaD$ is 
\be
{\bf {\hat B}} = \left( 
\begin{array}{ccc}
  0 & {\hat B}_{12} & 0 \\
  -{\hat B}_{12} & 0 & a_2 p_1 \\
  0 & -a_2 p_1 & 0
\end{array} \right) ,
\ee
where
\be
{\hat B}_{12} = [a_1 \sqrt{g_{11}} + \sqrt{6}(a_1{}^2 - a_2{}^2) g_{11}] p_1 .
\ee

\section{Four-dimensional generalizations}
We can obtain a four-dimensional spacetime simply by adding a time
coordinate according to the prescription
\be
^{(4)}ds^2 = -(dq^0)^2 + ds^2 ,
\ee
where $ds^2$ is a three-dimensional positive-definite metric. It
follows that
\be
^{(4)}{\bf \Gamma} = \left( 
\begin{array}{cc}
  0 & 0 \\
  0 & {\bf \Gamma}
\end{array} \right) .
\ee
For the cases obtained above this will lead to inequivalent spacetimes.
One way to generalize the three-dimensional Lax pair is
\be
^{(4)}{\bf L} = \left( 
\begin{array}{cc}
  i \, p_0 & 0 \\
  0 & {\bf L}
\end{array} \right) ,
\quad
^{(4)}{\bf A} = \left( 
\begin{array}{cc}
  0 & 0 \\
  0 & {\bf A}
\end{array} \right) ,
\ee
for which
\be
^{(4)}{\bf B} = \left( 
\begin{array}{cc}
  0 & 0 \\
  0 & {\bf B}
\end{array} \right) .
\ee
This Lax pair gives the geodesic Hamiltonian of the corresponding
spacetime metric as quadratic invariant.

\subsection{Case I}
Adding a time dimension to (\ref{eq:metric1}) we obtain the metric
\be\label{eq:4metric1}
ds^2 = -(dq^0)^2 + g_{11} (dq^1)^2 + (dq^2)^2 + g_{33} (dq^3)^2 ,
\ee
where
\be\eqalign{
  g_{11} &= \left(1 + 2 a_1{}^2 \right)^{-1}, \cr
  g_{33} &= \left(1 + 2 a_2{}^2 \right)^{-1}, \cr
  a_1 &= \exp(q^1 - q^2), \cr
  a_2 &= \exp(q^2 - q^3).}
\ee
This spacetime is of Petrov type I. The
nonzero components of the energy-momentum tensor calculated in a Lorentz frame
are ($\kappa=1$)
\be\eqalign{
  T^{00} &= -((g_{11})^2 \, T^{11} + T^{22} + (g_{33})^2 \, T^{33}) \cr
  &= -4 a_1{}^2 (g_{11})^2 (a_1{}^2 - 1) + 4 a_1{}^2 a_2{}^2 g_{11} g_{33} -
      4 a_2{}^2 (g_{33})^2 (a_2{}^2 - 1) , \cr
  T^{11} &= 4 a_1{}^2 (a_1{}^2 - 1) , \cr
  T^{22} &= -4 a_1{}^2 a_2{}^2 g_{11} g_{33} , \cr
  T^{33} &= 4 a_2{}^2 (a_2{}^2 - 1) . }
\ee 

\subsection{Case II}
Adding a time dimension to (\ref{eq:metric2}) we obtain the metric
\be\label{eq:4metric2}
ds^2 = -(dq^0)^2 + g_{11} (dq^1)^2 + (dq^2)^2 + (dq^3)^2 ,
\ee
where 
\be\eqalign{
  g_{11} &= \left[ 1 + 2(a_1{}^2 + a_2{}^2) \right]^{-1} , \cr
  a_1 &= \exp(\case1{\sqrt{2}} q^1 + \sqrt{\case32} q^2) , \cr
  a_2 &= \exp(\case1{\sqrt{2}} q^1 - \sqrt{\case32} q^2) . }
\ee
This spacetime is of Petrov type D. The nonzero components of the
energy-momentum tensor calculated in a Lorentz frame are ($\kappa=1$)
\be\eqalign{
  T^{00} &= 12 \, e^{2\sqrt{2}q^1} \, g_{11}{}^2
  \left(4 - 2 \sinh^2(\sqrt{6} q^2) +
    e^{-\sqrt{2} q^1} \, \cosh(\sqrt{6} q^2) \right) , \cr
  T^{11} &= -(g^{11})^2 \, T^{00} . }
\ee 

\subsection{Comment on the energy-momentum tensors}
In both of the above cases, the energy-momentum tensor takes the form
\be
T^{\alpha\beta} = \left( 
\begin{array}{cccc}
  \mu & 0 & 0 & 0 \\
  0 & p_1 & 0 & 0 \\
  0 & 0 & p_2 & 0 \\
  0 & 0 & 0 & p_3
\end{array} \right) ,
\ee
where $\mu := T^{00}$ is the energy density, and
$p_i := T^{ii}, (i = 1, 2, 3)$ are anisotropic pressures. Such an
energy-momentum tensor is physically meaningful if the weak
energy condition \cite{book:HawkingEllis1973}
\be\eqalign{
  \mu &\geq 0 , \cr
  \mu + p_i &\geq 0, \quad i =  1, 2, 3 , }
\ee
is satisfied. For case {\rm I}, there is an unbounded subdomain of the space
coordinates $(q^1, q^2, q^3)$ for which the weak energy condition holds.
For case {\rm II}, it is easily seen that the restrictions on the
energy-momentum tensor are inconsistent, so that the weak energy
condition never holds.

\section{Discussion}
In this paper we have presented the first application of the
tensorial Lax pair approach to integrable geometries.
Two inequivalent geometries representing the three-particle open Toda
lattice were found. This reflects the fact that the same underlying
mathematical structure may correspond to inequivalent physical
systems.

The geometrization procedure used in this work relies on 
canonical transformations which are peculiar to the particular problem
considered. Other more general geometrization schemes involving
reparameterizations of the independent variable may also be used to
construct integrable geometries. However it is not known at present
how to transform the Lax representation under time
reparameterizations.
This is despite the fact that it is known how to transform the
invariants themselves \cite{art:RosquistPucacco1995,art:Rosquist1997}.
The ans\"atze for the metrics (\ref{eq:4metric1}) and (\ref{eq:4metric2})
are not the most general one can think of. One possible generalization
is to include time-dependence in the metric coefficients. Furthermore,
by starting with other integrable systems, one would expect to obtain
new examples of integrable spacetimes. The possibility to find
physically interesting integrable geometries with a Lax pair is thus
not exhausted by the present work.

%\bibliography{lax}

\end{document}